\documentclass[aip,
nofootinbib,
rsi,%
reprint,
twocolumn,
author-numerical%
]{revtex4-1}

\usepackage{amsmath,amssymb,graphicx}
\usepackage[usenames]{color}

\usepackage{setspace}
\usepackage{lineno}

\newcommand{\pdr }[2]{\dfrac{\partial   {#1}}{\partial {#2}}}

\newcommand{\pdra }[2]{{\partial   {#1}}/{\partial {#2}}}

\newcommand{\tz  }{\tilde{z}}
\newcommand{\tv  }{\tilde{v}}

\newcommand{\tj  }{\tilde{j}}
\newcommand{\tJ  }{\tilde{J}}

\newcommand{\tR  }{\tilde{R}}

\newcommand{\teta}{\tilde{\eta}}

\newcommand{\tZ  }{\tilde{Z}}

\newcommand{\tc   }{\tilde{c}}

\newcommand{\flam}{\phi_{\lambda}}

\newcommand{\lexp}[1]{\exp\left(#1\right)}
\newcommand{\lnl }[1]{\ln \left(#1\right)}

\newcommand{\cref  }{c_{ref}}

\newcommand{\Rom   }{R_{\Omega}}

\newcommand{\lam   }{\lambda}

\newcommand{\lcat }{l_t}

\newcommand{\etal}{et{} al.{} }

\newcommand{\tit}{\tilde{t}}
\newcommand{\Cdl}{C_{dl}}

\newcommand{\ri}{{\rm i}}
\newcommand{\tom}{\tilde{\omega}}

\begin{document}

\sf

\title{Performance of a PEM fuel cell with oscillating air flow velocity: 
A modeling study based on cell impedance}

\author{Andrei Kulikovsky}
\thanks{ECS Active member}
\email{A.Kulikovsky@fz-juelich.de}

\affiliation{Forschungszentrum J\"ulich GmbH  \\
Institute of Energy and Climate Research,     \\
IEK--14: Electrochemical Process Engineering   \\
D--52425 J\"ulich, Germany
}
\altaffiliation[Also at:]{Lomonosov Moscow State University,
                Research Computing Center, 119991 Moscow, Russia}

\begin{abstract}
A model of PEM fuel cell impedance is developed taking into account imposed harmonic 
perturbation of the air flow velocity in the cathode channel. 
The flow velocity modulation with the amplitude 
proportional to AC amplitude of the cell potential lowers
the resistivity $R_h$ due to oxygen transport in channel.
When relative amplitudes of velocity and potential oscillations are equal,
a complete compensation of $R_h$ occurs. This effect explains 
experimental findings of Kim \etal 
(doi:10.1016/j.jpowsour.2008.06.069) and Hwang \etal 
(doi:10.1016/j.ijhydene.2010.01.064), who demonstrated significant improvement 
of PEM fuel cell performance under oscillating air flow velocity.
\end{abstract}

\maketitle


\section{Introduction}

PEM fuel cell needs air (oxygen) for protons and electrons conversion into water.
Air is usually supplied to the cell cathode through a system of channels. 
As any other transport process in the cell, oxygen transport through the channel 
is equivalent to electric resistivity $R_h$ leading to potential loss.
In more general terms, one has to speak about impedance $Z_h$ of oxygen transport
in the channel~\cite{Ingo_07a,Ingo_07b}.

In 2007, Schneider \etal\cite{Ingo_07a,Ingo_07b} have attracted attention 
of fuel cell community to  ``channel'' impedance, 
a ``forgotten player'' in the theory of PEMFC impedance. Since that time, a number of experimental~\cite{Reshetenko_11a,Reshetenko_13,Zamel_13b}
and modeling~\cite{Kulikovsky_12f,Bao_15,Kulikovsky_15g,Chevalier_16b,%
Kulikovsky_17f,Chevalier_18,Kulikovsky_18a} 
studies of this impedance have been published.
At typical air (oxygen) flow stoichiometry of about 2, the 
contribution of $R_h$ to the total cell resistivity is about 15\%~\cite{Kulikovsky_17f}.
Clearly, lowering of $R_h$ would lead to significant improvement of the cell performance.

Kim \etal\cite{Kim_08b} and later Hwang \etal\cite{Hwang_10} 
experimentally demonstrated dramatic improvement of PEMFC performance 
under oscillating air flow velocity in the channel. 
The effect of flow pulsation on the cell performance was more pronounced
at lower air flow rates, and the cell performance increased with the amplitude of velocity 
pulsation~\cite{Kim_08b}. The gain in performance has been attributed 
to improvement of oxygen transport 
through the cell due to flow pulsation\cite{Kim_08b,Hwang_10}.

Below, a model for PEMFC impedance operated under oscillating air flow velocity
is developed. We show that flow velocity oscillations lead to lowering
of oxygen transport impedance in the cathode channel. Under certain relation between velocity 
and potential oscillation amplitudes, the resistivity of oxygen transport in the channel
vanishes. This result supports the general conclusion 
of Kim \etal\cite{Kim_08b} and Hwang \etal\cite{Hwang_10}
that flow pulsation improves oxygen transport in the cell;
in this work, we demonstrate the mechanism of this improvement. Further,
the model gives a relation between the amplitudes 
of velocity and potential oscillations, at which the oxygen transport 
loss in the channel vanishes.

\section{Model}

\begin{figure}
\begin{center}
\includegraphics[viewport = 100pt 460pt 500pt 650pt, clip, scale=0.6]{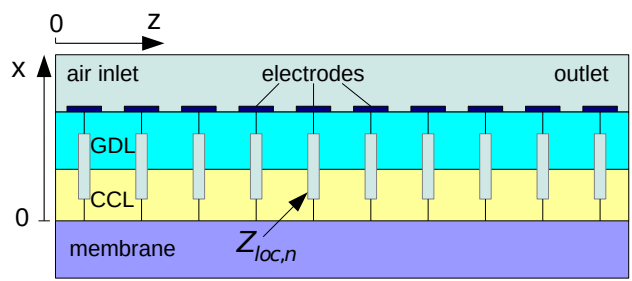}
\caption{Schematic of the segmented single--channel cell.
}
\label{fig:sketch}
\end{center}
\end{figure}

The model of PEMFC impedance below is extension of the model~\cite{Kulikovsky_19b}.
Consider a segmented PEM fuel cell equipped with the single straight cathode channel
(Figure~\ref{fig:sketch}). In the channel, oxygen is assumed to be transported 
along the $z$--axis, while in the porous layers it is transported 
in the through--plane direction to the cathode catalyst layer (CCL), where the 
oxygen reduction reaction (ORR) takes place.    
The characteristic frequency of oxygen transport in cathode channel is~\cite{Kulikovsky_19b}
\begin{equation}
   f_h \simeq \dfrac{3.3v}{2\pi L}
   \label{eq:fh}
\end{equation}
where 
$v$ is the air flow velocity and
$L$ is the channel length. For typical flow velocity on the order of $10^2$~cm~s$^{-1}$ 
and the channel length $L \simeq 100$~cm, we get $f_h \simeq 0.5$~Hz. 
This frequency is well below characteristic frequencies for 
the oxygen and proton transport in porous layers~\cite{Kulikovsky_19b}, 
and hence in the analysis of low--frequency phenomena
the latter processes can be ignored. 
The impedance model can thus be derived 
from the performance model, which takes into account oxygen transport
in the channel and faradaic process in the cell. 

\subsection{Performance model}

Assuming fast proton and oxygen transport in the trough--plane direction, 
the cell performance is described  by the oxygen mass transport equation
in the channel
\begin{equation}
   \pdr{c(t,z)}{t} + v(t)\pdr{c(t,z)}{z} = -\dfrac{j(z)}{4Fh}, \quad c(0) = \cref
   \label{eq:cz}
\end{equation}
and proton current conservation equation
\begin{equation}
   \Cdl\lcat\pdr{\eta(t)}{t} - j(z) = -\lcat i_* \left(\dfrac{c(t,z)}{\cref}\right)\lexp{\dfrac{\eta(t)}{b}}
   \label{eq:eta}
\end{equation}
Here,
$c(t,z)$ is the oxygen concentration concentration in the channel,
$\cref$ is the reference concentration,
$z$ is the distance along the channel, 
$j$ is the cell current density,
$h$ is the channel depth,
$\eta(t,z)$ is the ORR overpotential, positive by convention,
$\lcat$ is the CCL thickness,
$i_*$ is the ORR exchange current density, 
$b$ is the ORR Tafel slope.
     
Eq.\eqref{eq:cz} expresses oxygen mass balance assuming plug flow 
conditions  in the channel. The right side of this equation is the stoichiometric flow 
of oxygen through the membrane--electrode assembly, which agrees with the assumption of 
fast O$_2$ transport through the MEA. 
 
Eq.\eqref{eq:eta} is the proton charge conservation equation in the CCL, assuming 
that the rate of proton transport through the CCL is fast. 
This assumption means that the ORR overpotential $\eta$ 
is nearly constant through the CCL depth. 
The overpotential $\eta$ is also 
assumed to be independent of the distance $z$; this assumption holds if 
electron conductivity of the cell is large and ohmic losses in the cell
are small~\cite{Kulikovsky_19a}.  The first term on the 
left side of Eq.\eqref{eq:eta} describes the displacement current during 
charging/discharging of a double layer, and the term on the right side is the
local proton current consumed in the ORR. 

In this work, the flow velocity $v$ in Eq.\eqref{eq:cz} is considered as the time--dependent 
variable. 
With the dimensionless variables
\begin{multline}
   \tit = \dfrac{t}{t_*}, \quad \tz = \dfrac{z}{L}, \quad \tj = \dfrac{j}{i_*\lcat},
   \quad \tc = \dfrac{c}{\cref}, \\
   \teta = \dfrac{\eta}{b}, \quad \tv = \dfrac{v}{v_*}, 
   \quad \tZ = \dfrac{Z i_* \lcat}{b}, \quad \tom = \omega t_* 
   \label{eq:dless}   
\end{multline}
Eqs.\eqref{eq:cz} and \eqref{eq:eta} take the form
\begin{align}
      &\psi^2\pdr{\tc}{\tit} + \tv\lam\tJ\pdr{\tc}{\tz} = - \tj, \quad \tc(0) = 1 
      \label{eq:tcz} \\
      &\pdr{\teta}{\tit} - \tj = -\tc \exp\teta
      \label{eq:teta}
\end{align}
where $\psi$ is the dimensionless parameter
\begin{equation}
   \psi = \sqrt{\dfrac{4 F h \cref}{\Cdl b \lcat}},
   \label{eq:psi}
\end{equation}
$t_*$ is the characteristic time of the double layer charging  
\begin{equation}
   t_* = \dfrac{\Cdl b}{i_*},
   \label{eq:tast}
\end{equation}
$v_*$ is the time--average flow velocity (see below), 
$\lam$ is the stoichiometry of air flow corresponding to steady--state 
flow with the velocity $v_*$
\begin{equation}
   \lam  = \dfrac{4 F v_* h \cref}{L J}
   \label{eq:lam}
\end{equation}
and $\tJ$ is the mean current density in the cell
\begin{equation}
   \tJ = \int_0^1 \tj\, d\tz  .
   \label{eq:tJ}
\end{equation}
 
A key difference of the system \eqref{eq:tcz}, \eqref{eq:teta} 
from the system considered in~\cite{Kulikovsky_19b} 
is that $\tv$ in Eq.\eqref{eq:tcz} is a function of time.

\subsection{Impedance}

Now we apply small--amplitude harmonic perturbations to Eqs.\eqref{eq:tcz}, \eqref{eq:teta}.
The perturbations are of the form
\begin{equation}
   \begin{split}
   &\teta = \teta^0 + \teta^1\exp(\ri\tom\tit) \\
   &\tj   = \tj^0 + \tj^1\exp(\ri\tom\tit) \\
   &\tc   = \tc^0 + \tc^1\exp\exp(\ri\tom\tit)
   \end{split}
   \label{eq:Fourier}
\end{equation}
Assuming that the inlet flow velocity 
is modulated with the amplitude proportional to the amplitude of potential
perturbation, the time dependence of $\tv$ can be written as
\begin{equation}
   \tv   = 1 + k_v\teta^1\exp(\ri\tom\tit)
   \label{eq:tv01}  
\end{equation}
where $0 \leq k_v \leq 1$ is 
the real and non--negative modulation amplitude parameter. 
The unperturbed flow velocity is $v_*$ and hence
the static term in Eq.\eqref{eq:tv01} is unity. Note that in experiments 
of Hwang~\etal\cite{Hwang_10}, the mean flow velocity was zero. In this case,  
the static term in Eq.\eqref{eq:tv01} is zero and 
the flow velocity has to be scaled using speed of sound, for example.  
Note also that real $k_v$ means
that there is no phase shift between $\teta$ and $\tv$ oscillations;
these oscillations may differ only in amplitude.    

Substituting Eqs.\eqref{eq:Fourier} and \eqref{eq:tv01} 
into Eqs.\eqref{eq:tcz}, \eqref{eq:teta} and neglecting terms with 
the perturbation products, we get equations for the perturbation amplitudes
\begin{multline}
   \lam \tJ \pdr{\tc^1}{\tz} = -\left({\rm e}^{\teta^0} + \ri\tom\psi^2 \right)\tc^1 \\
   - \left({\rm e}^{\teta^0}\tc^0 + \ri\tom \right)\teta^1 
   - \lam\tJ\pdr{\tc^0}{\tz}k_v\teta^1, \quad   \tc^1(0) = 0
   \label{eq:tc1z}
\end{multline}
\begin{equation}
   \tj^1 = {\rm e}^{\teta^0}\left(\tc^1 + \tc^0\teta^1\right) + \ri\tom\teta^1
   \label{eq:tj1}
\end{equation}
where Eq.\eqref{eq:tc1z} is obtained using Eq.\eqref{eq:tj1}. The boundary condition
to Eq.\eqref{eq:tc1z} means that the inlet oxygen concentration is not perturbed; 
perturbed is the flow velocity only. Generally, if the flow velocity is perturbed using 
pressure modulation, the inlet oxygen concentration would also oscillate and the boundary 
condition to Eq.\eqref{eq:tc1z} would read $\tc^1(0) = \tc^1_0$. This 
condition, however, complicates the analysis not changing the main results. 

The goal of this work is to demonstrate the effect of inlet velocity modulation 
on the cell impedance and for the shapes of static current and oxygen 
concentration along the channel we take the zero--order solutions~\cite{Kulikovsky_19a}:
\begin{align}
   &\tj^0 = -\tJ\lam\lnl{1 - \dfrac{1}{\lam}}\left(1 - \dfrac{1}{\lam}\right)^{\tz} 
   \label{eq:tj0z} \\
   &\tc^0 = \left(1 - \dfrac{1}{\lam}\right)^{\tz} 
   \label{eq:tc0z}
\end{align} 
Eqs.\eqref{eq:tj0z}, \eqref{eq:tc0z} are valid if the cell ohmic resistivity $\Rom$ is small, 
i.e., the product  $J \Rom/b \ll 1$ (see~\cite{Kulikovsky_19a} for details).
Eqs.\eqref{eq:tj0z}, \eqref{eq:tc0z} allow us to get analytical solution to the problem.
A more accurate approximation of $\tj^0$ and $\tc^0$ can be obtained numerically as 
discussed in~\cite{Kulikovsky_19a}. 

Local cell impedance at the point $\tz$ is given by
\begin{equation}
   \tZ_{loc}(\tz) = \dfrac{\teta^1}{\tj^1}
   \label{eq:tZlocdef}
\end{equation}

To calculate $\tZ_{loc}$, we solve Eq.\eqref{eq:tc1z}:
\begin{multline}
   \tc^1 = \dfrac{\ri\teta^1 \tom \left(\left(1 - \dfrac{1}{\lam}\right)^{\tz}
                \lexp{-\dfrac{\ri\tom \psi^2\tz}{\lam \tJ}} - 1\right)}
                             {\flam\tJ + \ri\tom \psi^2} \\
   + \dfrac{\teta^1 \left(1 -  k_v \right)\flam \tJ}{\ri\tom\psi^2}\left(1 - \dfrac{1}{\lam}\right)^{\tz} 
     \left(\lexp{-\dfrac{\ri\tom \psi^2\tz}{\lam \tJ}} - 1\right) 
   \label{eq:tc1sol}
\end{multline}
where the parameter $\flam$ is 
\begin{equation}
    \flam = -\lam\lnl{1 - \dfrac{1}{\lam}},
   \label{eq:flam}
\end{equation}
and equation for the static cell polarization curve
\begin{equation}
   \flam\tJ = {\rm e}^{\teta^0}
   \label{eq:vcc}
\end{equation}
was used to eliminate ${\rm e}^{\teta^0}$ in Eq.\eqref{eq:tc1sol}.
Eq.\eqref{eq:vcc} is obtained upon substitution of Eqs.\eqref{eq:tj0z}, \eqref{eq:tc0z} into
the static version of charge conservation equation \eqref{eq:teta}.

Substituting Eq.\eqref{eq:tc1sol}
into Eq.\eqref{eq:tj1} and dividing the resulting equation by $\tj^1$, we get an algebraic equation 
for $\tZ_{loc}$. Solving this equation, we come to 
\begin{multline}
   \tZ_{loc} = \dfrac{1}{\flam\tJ}\Biggl\{\ri\left(\dfrac{\tom}{\flam\tJ + \ri\tom \psi^2} - \dfrac{\flam\tJ (1 - k_v)}{ \tom\psi^2}\right) \\
   \times\left(1 - \dfrac{1}{\lam}\right)^{\tz} \lexp{-\dfrac{\ri\tom \psi^2\tz}{\lam \tJ}} 
   - \dfrac{\ri\tom}{\flam\tJ + \ri\tom \psi^2} \\
   + \left(1 - \dfrac{1}{\lam}\right)^{\tz}\left(1 + \dfrac{\ri\flam\tJ (1 - k_v)}{ \tom\psi^2}\right)
   + \dfrac{\ri\tom}{\flam\tJ}\Biggr\}^{-1}
   \label{eq:tZloc}
\end{multline}

The total cell impedance $\tZ_{cell}$ is given by
\begin{equation}
   \tZ_{cell} = \left(\int_0^1\dfrac{d\tz}{\tZ_{loc}}\right)^{-1}
   \label{eq:tZcell}
\end{equation}
Calculation of integral gives
\begin{equation}
   \tZ_{cell} = \left(\left(\ri\flam\tJ - 2\tom\psi^2\right)\flam\tJ 
                           - \ri\tom^2 \psi^4\right)\dfrac{\tom\psi^2}{D_{cell}}
   \label{eq:tZcell_res}
\end{equation}
where
\begin{multline}
   D_{cell} =  (\lam - 1) \left(1 - \lexp{-\dfrac{\ri \tom \psi^2}{\lam \tJ}}\right)(1 - k_v)\flam^3\tJ^4 \\
   - \left( (1 - k_v)\left(\lexp{-\dfrac{\ri \tom \psi^2}{\lam \tJ}} (\lam - 1) - \lam\right)  - 2 k_v + 1\right) \\
   \times\ri \tom\psi^2\flam^2 \tJ^3 \\
   - \left((1 + k_v) \psi^2 - \lexp{-\dfrac{\ri \tom \psi^2}{\lam \tJ}} (\lam - 1) + \lam\right) \\
  \times \tom^2 \psi^2\flam \tJ^2  
   - \ri\left(\psi^2 + \flam\right)\tom^3\psi^4\tJ + \tom^4 \psi^6
\end{multline}

\section{Results and discussion}

\begin{figure}
\begin{center}
\includegraphics[viewport = 80pt 530pt 320pt 700pt, clip, scale=1]{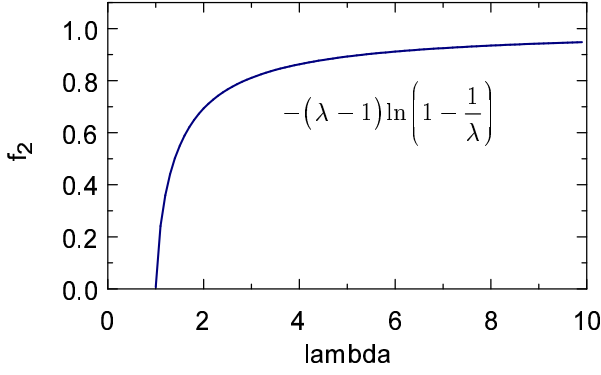}
\caption{Function $\phi_2 = - (\lam - 1)\lnl{1 - 1/\lam}$ appearing in
denominator of Eqs.\eqref{eq:Rcell},\eqref{eq:Rcell0}. 
}
\label{fig:f2}
\end{center}
\end{figure}

It is advisable to consider first the limit of $\omega \to 0$.
Expansion of Eq.\eqref{eq:tZcell_res} in Taylor series over $\tom \to 0$ gives at leading order 
the differential cell resistivity $\tR_{cell}$, which in dimension form is
\begin{equation}
   R_{cell} = \dfrac{b}{J \biggl(k_v - (\lam - 1)\lnl{1 - 1/\lam}(1 - k_v)\biggr)}
   \label{eq:Rcell}
\end{equation}
With $k_v =0$ (no velocity modulation), Eq.\eqref{eq:Rcell} reduces to~\cite{Kulikovsky_19b}
\begin{equation}
   R_{cell}^{k_v=0} = - \dfrac{b}{J \biggl((\lam - 1)\lnl{1 - 1/\lam}\biggr)}
   \label{eq:Rcell0}
\end{equation}
The factor 
\begin{equation}
   \phi_2 = (\lam - 1)\lnl{1 - \dfrac{1}{\lam}} < 1
   \label{eq:f2}
\end{equation}
in denominator of Eq.\ref{eq:Rcell0} describes the resistivity growth
due to finite air flow stoichiometry $\lam$; the 
less is $\phi_2$, the larger is the transport resistivity  (Figure~\ref{fig:f2}). 

However, with $k_v=1$ the 
dependence on $\lam$ in Eq.\eqref{eq:Rcell} vanishes and we get
\begin{equation}
   R_{cell}^{k_v=1} = \dfrac{b}{J}
   \label{eq:Rcell1}
\end{equation}
which is a pure charge--transfer cell resistivity~\cite{Kulikovsky_15a}. 
Thus, velocity perturbation with the dimensionless
amplitude equal to the amplitude of potential perturbation completely 
compensates for the losses due to oxygen transport in the cathode channel.
The equality of $\tv^1$ and $\teta^1$ perturbation amplitudes
means that the following relation between the dimension amplitudes
must hold:
\begin{equation}
   \dfrac{v^1}{v_*} = \dfrac{\eta^1}{b}
   \label{eq:veta}
\end{equation} 
Typical ORR Tafel slope in Pt/C electrodes is about 30~mV. Thus, 
with the potential oscillation amplitude on the order of 3~mV, the flow velocity
oscillation with the amplitude of 10\% 
of the time--average velocity provides complete 
compensation of oxygen transport losses in the channel.   

With the growth of $\lam$, the effect of velocity modulation
progressively lowers. Indeed, setting in Eq.\eqref{eq:Rcell1} $\phi_2 \simeq 1$,
we see that the dependence on $k_v$ vanishes. From Figure~\ref{fig:f2} is is clear
that the effect of velocity modulation is most pronounced at $\lam \lesssim 2$,
which agrees with the experimental results of Kim \etal\cite{Kim_08b}.

\begin{table}
\small
	\begin{center}
		\begin{tabular}{|l|c|}
			\hline
			Catalyst layer thickness $\lcat$, cm                          &   $10^{-3}$            \\
			Exchange current density $i_*$, A~cm$^{-3}$                   &   $10^{-3}$            \\
			ORR Tafel slope $b$, V                                        &   0.03                 \\
			Double layer capacitance, $\Cdl$, F~cm$^{-3}$                 &   20                   \\
			Channel depth $h$, cm                                         &   0.1                  \\
			Cell temperature  $T$, K                                      &   $273 + 80$           \\
			Mean cell current density $J$, A~cm$^{-2}$                    &   0.1                  \\
            Air flow stoichiometry $\lam$                                 &   2                    \\
			\hline
		\end{tabular}
	\end{center}
	\caption{Cell geometrical and operating parameters used in the calculations.
        The characteristic values of $b$ and $\Cdl$ are taken from impedance
        measurements~\cite{Kulikovsky_16b}; the value of $i_*$ is assumed.
        The Tafel slope is given per exponential basis.
	}
	\label{tab:parms}
\end{table}
%

\begin{figure}
\begin{center}
\includegraphics[viewport = 60pt 370pt 300pt 680pt, clip, scale=0.9]{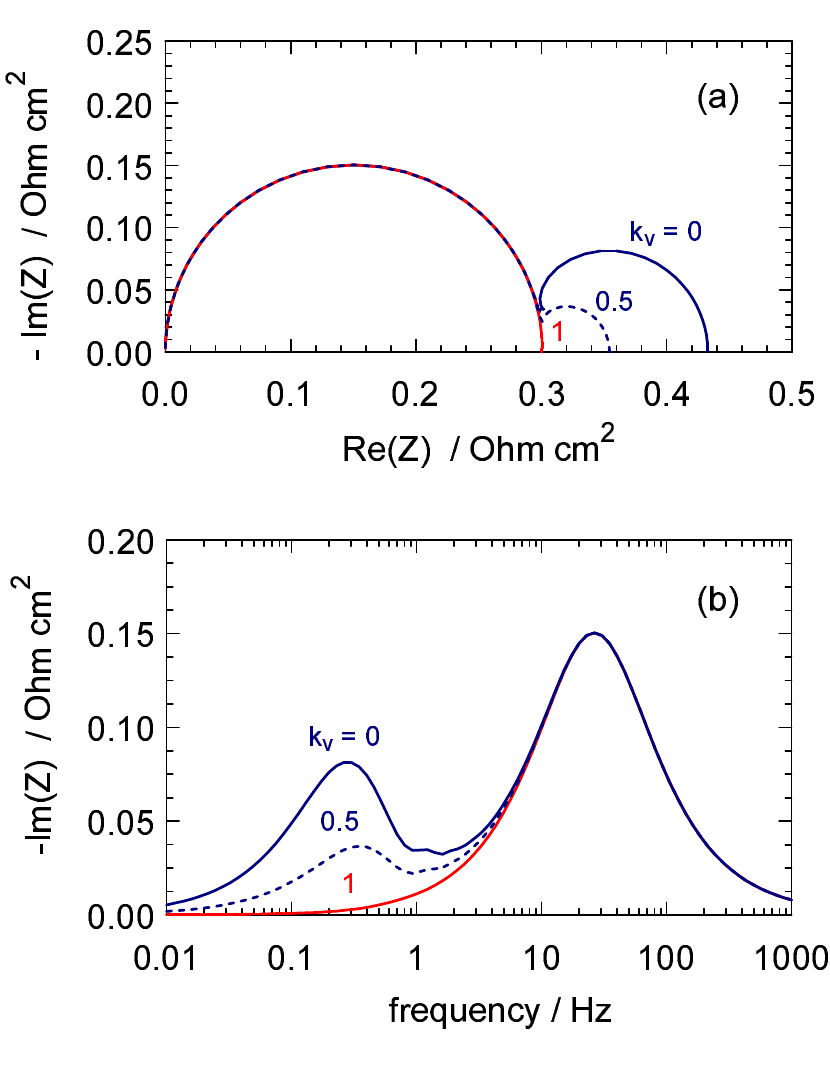}
\caption{(a) Nyquist spectra of the total cell impedance, Eq.\eqref{eq:tZcell_res} for 
   the indicated values of the flow velocity amplitude parameter $k_v$, Eq.\eqref{eq:tv01}.
   Zero $k_v$ corresponds to static flow velocity.
   (b) The frequency dependence of imaginary part of impedance in (a). 
}
\label{fig:Zcell}
\end{center}
\end{figure}

The effect of $k_v$ on the dimension Nyquist spectra of Eq.\eqref{eq:tZcell} 
is shown in Figure~\ref{fig:Zcell}.
With $k_v=0$, the spectrum has the form of two arcs, with the left arc due to 
faradaic impedance, and the right arc due to oxygen transport in channel~\cite{Ingo_07a}.
When $k_v$ varies from 0 to 1, the ``channel'' arc gets smaller, and at $k_v=1$
this arc almost completely vanishes (Figure~\ref{fig:Zcell}). 
The curve $k_v=1$ illustrates compensation of the ``channel'' losses 
by the applied flow velocity oscillations.

To understand the role of velocity oscillations, consider Eq.\eqref{eq:tj1}.
With $\tc^1=0$ (no perturbation of the oxygen concentration in channel),
this equation simplifies to
\begin{equation}
   \tj^1 = \left({\rm e}^{\teta^0}\tc^0 + \ri\tom\right)\teta^1
   \label{eq:tj1b}
\end{equation}
and hence the local cell impedance $\tZ_{loc} = \teta^1/\tj^1$ reduces to 
impedance of a parallel $RC$--circuit:
\begin{equation}
   \tZ_{loc}^{\tc^1=0} = \dfrac{1}{{\rm e}^{\teta^0}\tc^0 + \ri\tom} 
                       = \dfrac{1}{\tj^0 + \ri\tom} 
   \label{eq:tZRC}
\end{equation}
Using here $\tj^0$ from Eq.\eqref{eq:tj0z}, and calculating the total cell
impedance according to Eq.\eqref{eq:tZcell}, we get
\begin{equation}
   \tZ_{cell}^{\tc^1=0} = \dfrac{1}{\tJ + \ri\tom}
   \label{eq:tZcell0}
\end{equation}
which is pure charge--transfer impedance. 
Thus, the oxygen transport losses are represented by the term with $\tc^1$ in 
Eq.\eqref{eq:tj1}. The trick is that with $k_v=1$ this term is strongly damped.

\begin{figure}
\begin{center}
\includegraphics[viewport = 70pt 530pt 320pt 740pt, clip, scale=0.9]{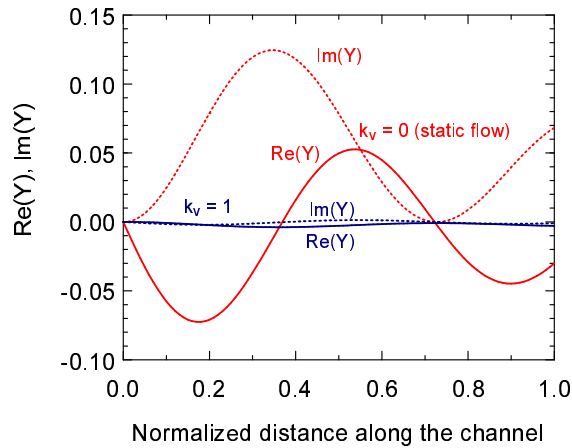}
\caption{Normalized oxygen concentration perturbation amplitude $Y=\tc^1/\teta^1$, 
Eqs.\eqref{eq:Yc}, \eqref{eq:tc1sol}, along the cathode channel 
for the two values of the flow velocity 
amplitude parameter $k_v$, Eq.\eqref{eq:tv01}. 
The frequency of potential and velocity oscillations $f = 1$~Hz. 
}
\label{fig:Yc}
\end{center}
\end{figure}

Figure~\ref{fig:Yc} shows the real and imaginary part of the normalized 
oxygen ``concentration admittance'' 
\begin{equation}
   Y = \dfrac{\tc^1}{\teta^1}
   \label{eq:Yc}
\end{equation} 
obtained from Eq.\eqref{eq:tc1sol} with $k_v=0$ and $k_v=1$. As can be seen, at
$k_v=1$ the amplitude of $\tc^1$ oscillations is strongly damped, leading to much lower
transport loss.

%

Note that with $k_v=1$, $Y$ is still non--zero at $\tom > 0$ (Figure~\ref{fig:Yc}), meaning
that complete compensation of the channel impedance occurs in the limit of $\tom \to 0$
only, while at a finite $\tom$, the cell impedance slightly differs from the faradaic
impedance, Eq.\eqref{eq:tZcell0}.

From this analysis it follows that the whole effect of oxygen transport loss
in channel is purely dynamic in nature. In the true steady state, finite 
oxygen stoichiometry 
$\lam$ only shifts the polarization curve as a whole along the potential
axis, not changing the slope of the curve (cell resistivity). 
Indeed, from the static polarization curve, Eq.\eqref{eq:vcc}, it follows that 
the true static differential cell resistivity 
$R_{cell}^0 = \pdra{\teta^0}{\tJ} = 1/\tJ$, which is a pure faradaic resistivity
independent of $\lam$. This result also follows from Eq.\eqref{eq:tZcell0}
However, small perturbations of flow parameters immediately lead to small oscillations
of oxygen concentration in channel. These oscillations, in turn, 
induce small oscillations of the cell potential, and the system enters 
the dynamic mode with the quasi--static resistivity given by Eq.\eqref{eq:Rcell0}.
Harmonic modulation of the flow velocity with $0 < k_v \leq 1$ allows one 
to lower this resistivity, as Eq.\eqref{eq:Rcell} shows.     
      
In reality, fuel cell never works in a true steady state; 
due to small variation of operating conditions and aging of cell components,
even in stationary experiments and applications the cell potential slowly 
varies with time. 
This variation corresponds to a small but nonzero $\tom > 0$, making
the ``channel'' resistivity quite significant. 
In automotive applications, fuel cells operate in intrinsically transient 
regimes and the cell voltage strongly varies with time. 

In this work, AC perturbation  $\teta^1$  of the cell potential and 
the velocity oscillation amplitude $k_v$ are assumed to be independent parameters. 
However, in real applications, the amplitude of flow velocity oscillations 
could be regulated by flow controller, while the respective potential 
perturbation would be a dependent, uncontrolled parameter. 
The relation between oscillation amplitudes of velocity
and cell potential in this case could be controlled experimentally. 
Another option would be excitation of flow velocity oscillations 
by pressure wave applied to the inlet flow. However, 
development of impedance model which would 
describe this situation is a much more challenging task. 
The experiments~\cite{Kim_08,Hwang_10} and the simple model above suggest that 
the problem deserves further studies.

\section{Conclusions}

The model of PEM fuel cell impedance is developed taking into account
air flow velocity oscillations applied in--phase with the AC potential 
perturbation. The model is based on oxygen mass transport equation 
in the cathode channel coupled to the proton current conservation equation
in the cathode catalyst layer. The model aims at description of 
low--frequency phenomena in the cell and it ignores proton and oxygen 
transport in the porous layers, assuming that this transport is fast.    

The model shows that velocity oscillations reduce 
the resistivity $R_h$ of oxygen transport in the cathode channel.
If the relative amplitudes of velocity and potential oscillations 
are equal, the resistivity $R_h$ vanishes. These results explain experimental 
findings of Kim \etal\cite{Kim_08b} and Hwang \etal\cite{Hwang_10}
who demonstrated dramatic improvement of PEM fuel cell performance
under oscillating air flow velocity.




\small
\vspace*{2em}
{\large\bf Nomenclature\\[0.5em]}

\begin{tabular}{ll}
	$\tilde{}$   &  Marks dimensionless variables                             \\
	$b$          &  ORR Tafel slope, V                                        \\
	$\Cdl$       &  Double layer volumetric capacitance, F~cm$^{-3}$          \\
	$c$          &  Oxygen molar concentration, mol~cm$^{-3}$      \\
	$\cref$      &  Reference oxygen concentration                \\
	             &  (at the channel inlet), mol~cm$^{-3}$      \\
	$F$          &  Faraday constant, C~mol$^{-1}$                            \\
	$f$          &  Regular  frequency, Hz                                    \\
	$J$          &  Mean cell current density, A~cm$^{-2}$                    \\
	$j$          &  Local cell current density, A~cm$^{-2}$                   \\
	$h$          &  Channel depth, cm                                         \\
	$\ri$        &  Imaginary unit                                            \\
	$i_*$        &  Volumetric exchange current density, A~cm$^{-3}$          \\
	$L$          &  Channel length,cm                                         \\
	$\lcat$      &  Catalyst layer thickness, cm                              \\
	$t$          &  Time, s                                                   \\
	$v$          &  Flow velocity in the cathode channel, cm~s$^{-1}$         \\
	$x$          &  Coordinate through the cell, cm                     \\
	$Z$          &  Impedance,  $\Omega$~cm$^2$                      \\
	$z$          &  Coordinate along the air channel, cm                 \\[1em]
\end{tabular}


{\bf Subscripts:\\}

\begin{tabular}{ll}

	$h$      & Air channel            \\ 
	$loc$    & Local impedance        \\
	$*$      & Characteristic or time--average value   \\[1em]

\end{tabular}


{\bf Superscripts:\\}

\begin{tabular}{ll} 
	$0$      & Steady--state value            \\
	$1$      & Small--amplitude perturbation  \\[1em]
\end{tabular}


{\bf Greek:\\}

\begin{tabular}{ll}
	$\lam$              &  Air flow stoichiometry     \\
    $\flam$              &  Dimensionless parameter, Eq.\eqref{eq:flam} \\
    $\psi$              & Dimensionless parameter, Eq.\eqref{eq:psi}  \\
	$\omega$            &  Angular frequency ($\omega = 2\pi f$), s$^{-1}$    \\
\end{tabular}

\end{document}